\documentclass[submission, Proceedings]{SciPost}

\hypersetup{
    colorlinks,
    linkcolor={red!50!black},
    citecolor={blue!50!black},
    urlcolor={blue!80!black}
}
\usepackage[bitstream-charter]{mathdesign}
\DeclareSymbolFont{usualmathcal}{OMS}{cmsy}{m}{n}
\DeclareSymbolFontAlphabet{\mathcal}{usualmathcal}

\begin{document}

\begin{center}
{\Large \textbf{The KM3NeT infrastructure: status and first results}}
\end{center}

\begin{center}
Annarita Margiotta\\ on behalf of the KM3NeT Collaboration (https://www.km3net.org/)
\end{center}

\begin{center}
INFN - Sez. Bologna and Dipartimento di Fisica e Astronomia - Alma Mater Studiorum, Universit\`a di Bologna, Italy 
\\
annarita.margiotta@unibo.it

\end{center}
\begin{center}
\today
\end{center}


\definecolor{palegray}{gray}{0.95}
\begin{center}
\colorbox{palegray}{
  \begin{tabular}{rr}
  \begin{minipage}{0.1\textwidth}
    \includegraphics[width=30mm]{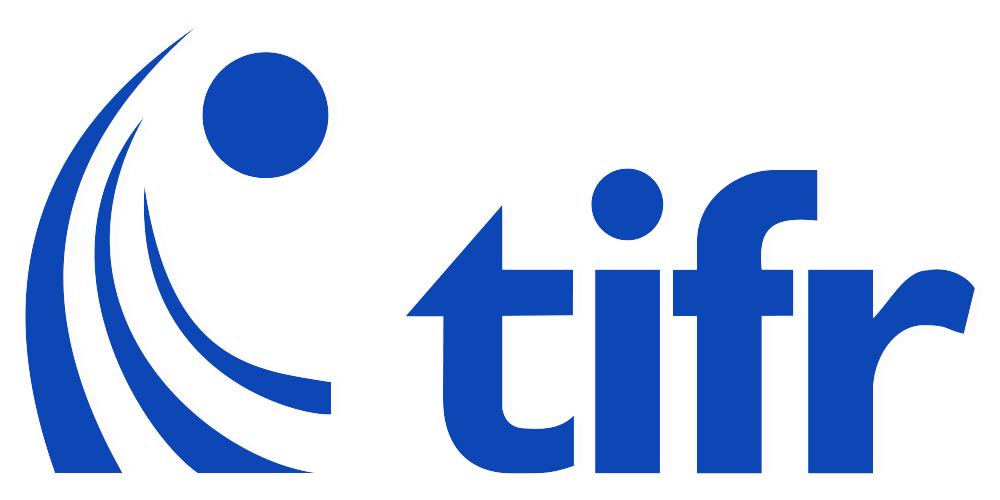}
  \end{minipage}
  &
  \begin{minipage}{0.85\textwidth}
    \begin{center}
    {\it 21st International Symposium on Very High Energy Cosmic Ray Interactions (ISVHECRI 2022)}\\
    {\it Online, 23-27 May 2022} \\
    \doi{10.21468/SciPostPhysProc.?}\\
    \end{center}
  \end{minipage}
\end{tabular}
}
\end{center}

\section*{Abstract}
{\bf
KM3NeT is a research infrastructure in construction under the Mediterranean Sea. It hosts two large volume neutrino Cherenkov telescopes: ARCA at a depth of 3500 m, located offshore Sicily, and ORCA, 2500 m under the sea level, offshore the southern French coast. The two detectors share the same detection principle and technology and the same data acquisition design, the only difference being the geometrical arrangement of the optical sensors.  This allows to span a wide range of neutrino energy and cover a large scientific program: the study of neutrino properties, first of all neutrino mass ordering, the identification and study of high energy neutrino  astrophysical sources, indirect dark matter searches and core collapse supernovae detection.}
\section{Introduction}
\label{sec:intro}
The KM3NeT collaboration is building a network of undersea neutrino detectors, which exploit the faint Cherenkov light produced along the pattern of relativistic charged particles emerging from the interactions of neutrinos with rock or water inside or in proximity of the detector to reconstruct the properties of the parent neutrino: ARCA (Astroparticle Research with Cosmics in the Abyss)   and ORCA (Oscillation Research with Cosmics in the Abyss).  The KM3NeT scientific program is very wide and ambitious and includes astrophysical items like the search for cosmic high energy neutrinos and the identification of their sources, indirect searches for dark matter,  detection of core collapse supernovae, and particle physics topics like the investigation of neutrino properties, in particular  the definition of neutrino mass ordering (NMO) through the measurement of matter effects in oscillation patterns of atmospheric neutrinos. To accomplish these goals neutrino detectors sensitive to neutrinos of different energy are required. In the following sections the main features of the KM3NeT telescopes,  their present status and expected performances are  described together with  some preliminary results, \cite {loi}.
\section{The KM3NeT neutrino telescopes: ARCA and ORCA}
\label{sec:km3net}
\begin{figure}[h]
\includegraphics[width=0.41\textwidth]{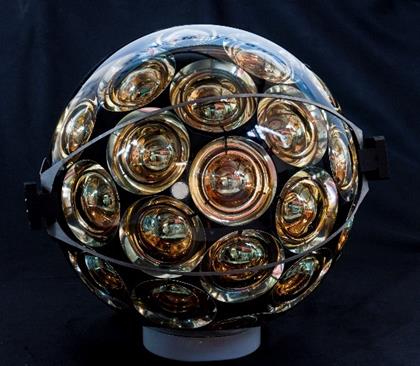}
\hfill
\includegraphics[width=0.625\textwidth]{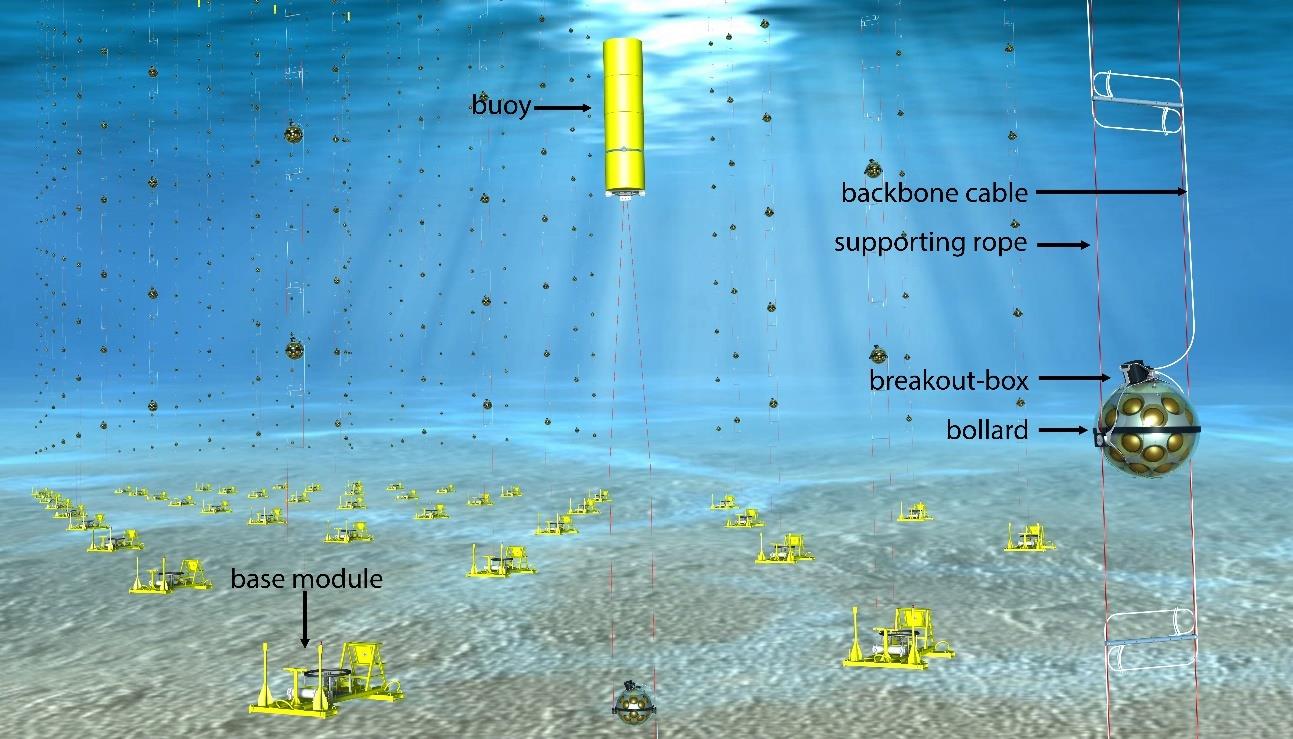}
\caption{\label{fig:DOM}Left: View of a KM3NeT DOM. Right: Artistic view of the KM3NeT telescope. Detection Units equipped with DOMs and anchored at the sea bed are shown. Buoys at the top of each string keep the structure taut. At the level of each DOM a breakout of the  backbone cable is visible. The Base Modules of DUs are connected to a Junction Box via cables deployed on the seafloor. The picture is not in scale.     }
\end{figure}

The KM3NeT neutrino telescopes are 3-dimensional lattices of optical sensors immersed in deep sea water. The optical sensor, called Digital Optical Module (DOM) \cite{dom}, is a glass sphere hosting 31 3" photomultipliers \cite{pmt}, Fig. {\ref{fig:DOM}}, left. Its design allows a better rejection of environmental background and improves track reconstruction \cite{proto}. The  light signals collected by  the  PMTs are digitised by  custom front-end electronic boards \cite{frontend} and sent to a computing farm located on-shore. According to the all-data-to-shore approach, all signals above a 0.3 photoelectron threshold are transmitted and processed with dedicated trigger algorithms looking for potentially interesting events \cite {cudaq}.
A set of 18 DOMs, distributed along a slender string and connected through a backbone cable and two  ropes, constitutes a Detection Unit (DU). Each DU is completed with a base module  containing electro-optical circuits that guarantee signal transmission between the string and the control stations located off-shore. A group of 115 DUs is called a Building Block (BB).  As of today 19 DUs are in data taking at the ARCA site and 10 DUs at the ORCA site. Completion of the infrastructure is foreseen in 2025 for ORCA and 2027 for ARCA. In autumn 2022 further sea operations during which more DUs will be deployed are scheduled. Fig. {\ref{fig:DOM}}, right, shows an artistic representation, not in scale, of the final detector. Technical and construction details  can be found in several KM3NeT technical papers here: https://www.km3net.org/about-km3net/publications/km3net-technical-publications/

Depending on the neutrino flavour and type of interaction, different signatures of events can be registered. Track-like events are mainly due to charged current muon neutrino interactions and are characterised by a long muon track. They represent the golden channel for the identification of neutrino sources because the parent neutrino direction can be reconstructed with an angular resolution of about 0.1° for neutrino energies above 100 TeV. The interaction vertex can be external to the detector and the energy of the event is calculated with a resolution of $\sim 0.3$ units in $\log_{10}(E_\mu)$.
Shower-like events are the result of neutral current interactions of all-flavour neutrinos and of charged current electron neutrino interactions. The direction of parent neutrino can be evaluated with a resolution better than a few degrees, while the energy is reconstructed with a precision better than 5 \%. The most abundant signal in this kind of detectors comes from the flux of atmospheric muons, \cite{O1A2}. In order to reduce this background a geometrical cut is applied and only upward going muons are considered. Atmospheric muons are  absorbed by the Earth and only neutrinos can produce  upward moving tracks. The bulk of this flux is due to atmospheric neutrinos that represent an irreducible background in searches for cosmic neutrinos, and the main signal for oscillation studies.  Only with accurate statistical analyses a cosmic flux can be identified, either looking for clusters of events over the atmospheric  background, or searching a neutrino emission from potential interesting sources, or looking for  a very high energy neutrino diffuse flux, above 50-100 TeV, where the conventional atmospheric neutrino flux is overwhelmed by the cosmic flux.

\subsection{ARCA} 
ARCA is being deployed at 3500 m below  sea level offshore Sicily. It will be made by two Building Blocks of Detection Units, 230 DUs in total, and will instrument about 1 km$^3$ of sea water thanks to a sparse array of DOMs, spaced by 36 m in vertical and about 90 m in horizontal. Its design is optimized  to look for high energy neutrinos in the energy range between 100 GeV and 100 PeV.  
\begin{figure}[h]
\includegraphics[width=0.47\textwidth]{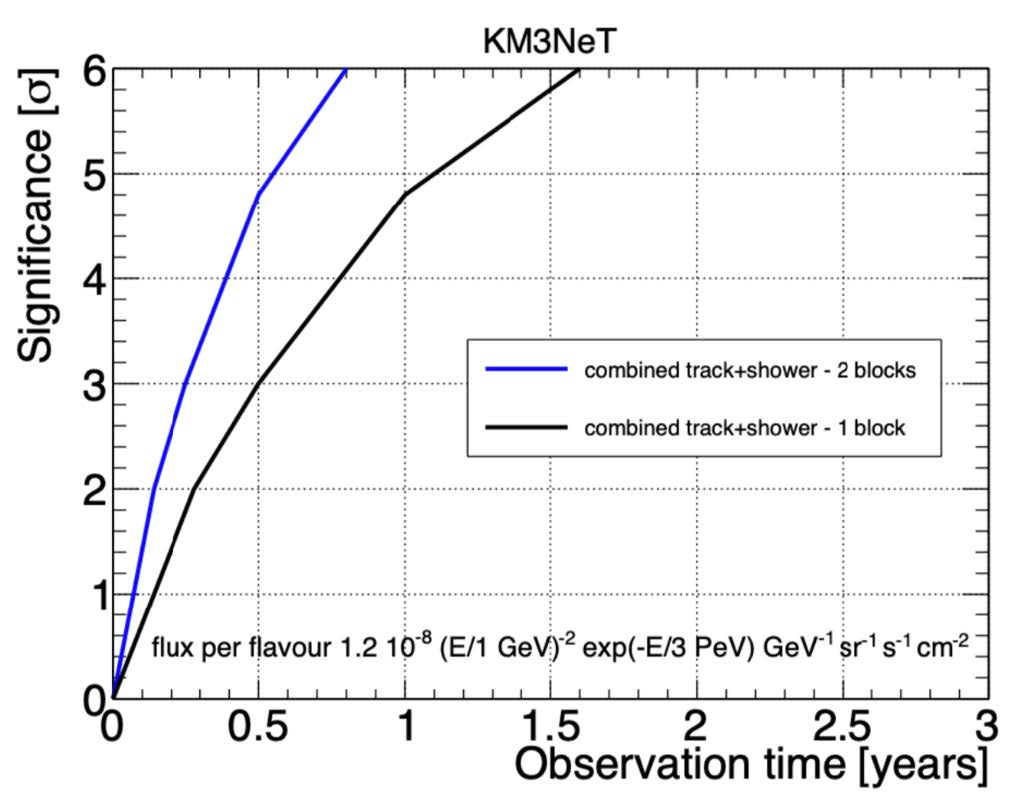}
\hfill
\includegraphics[width=0.54\textwidth]{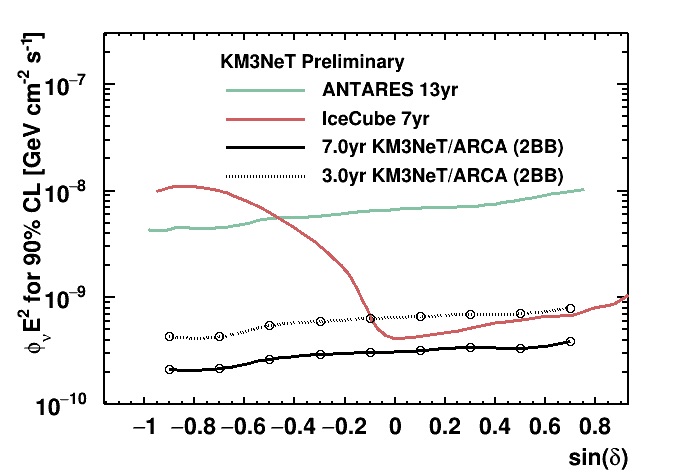}
\caption{Left: Significance for the detection of a diffuse flux of neutrinos as claimed by IceCube with 1 BB  (black line) or 2 BBs (blue line) of ARCA. Right: Sensitivity of ARCA (2 BB) to an $E^{-2}$ flux of neutrinos from point sources as a function of source declination compared to ANTARES and IceCube. }
\label{fig:significance}
\end{figure}
At present KM3NeT has secured funding for the construction of the first  Building Block of ARCA and part of the second BB.  The main goals of ARCA are  identification of Galactic and extra-Galactic  sources of neutrinos, where also extremely energetic cosmic rays might be produced and accelerated, and  measurement of a diffuse flux of neutrinos like the one detected by IceCube \cite{icecube}. As shown in Fig. {\ref{fig:significance}} (left), this latter result can be reached in about one year with a single BB and less than six months with two BBs. Fig. {\ref{fig:significance}} (right) shows ARCA sensitivity to neutrinos from point sources, assuming an $E^{-2} $ flux. Thanks to its size, to its location in the Northern hemisphere and to the optical properties of water, ARCA has a better sensitivity than present telescopes by more than one order of magnitude in the region of negative declination.

The scientific field of interest to ARCA also includes indirect searches for dark matter and the detection of core collapse supernovae. KM3NeT/ARCA is  involved in an intense multimessenger program to search for neutrinos emitted in space-time coincidence  with other astrophysical signals, like gravitational waves, high energy cosmic rays and photons over a wide energy range. 
\subsection{ORCA}
ORCA is a neutrino telescope in construction in front of the southern French coast, at 2500 m below the sea level. The detection principle and the general design of the detector is identical to ARCA's, but the denser distribution of the optical sensors, spaced by about 20 meters in horizontal and 9 meters in vertical, along the Detection Units, is optimized for the detection of neutrinos in the energy range 1-100 GeV. This  feature will allow to compare atmospheric neutrino oscillation patterns, in order to study the NMO and investigate in detail neutrino oscillation parameters, \cite{oscillation}. Sensitivity of ORCA to the identification of the NMO after 3 years of data taking is shown in Fig. {\ref{fig:nmo}.   In Fig. {\ref{fig:oscill} (left panel), the oscillation pattern measured with  six ORCA DUs in almost one year of data taking is shown, compared to the no-oscillation hypothesis and to the NuFit5.0 evaluation  \cite{nufit} assuming normal mass ordering of neutrinos. In the right panel  measurements of several other experiments are compared to the ORCA6 results. The purple line indicates the expectation with 1 BB of ORCA.
\begin{figure}[h]
\centering
\includegraphics[width=0.5\textwidth]{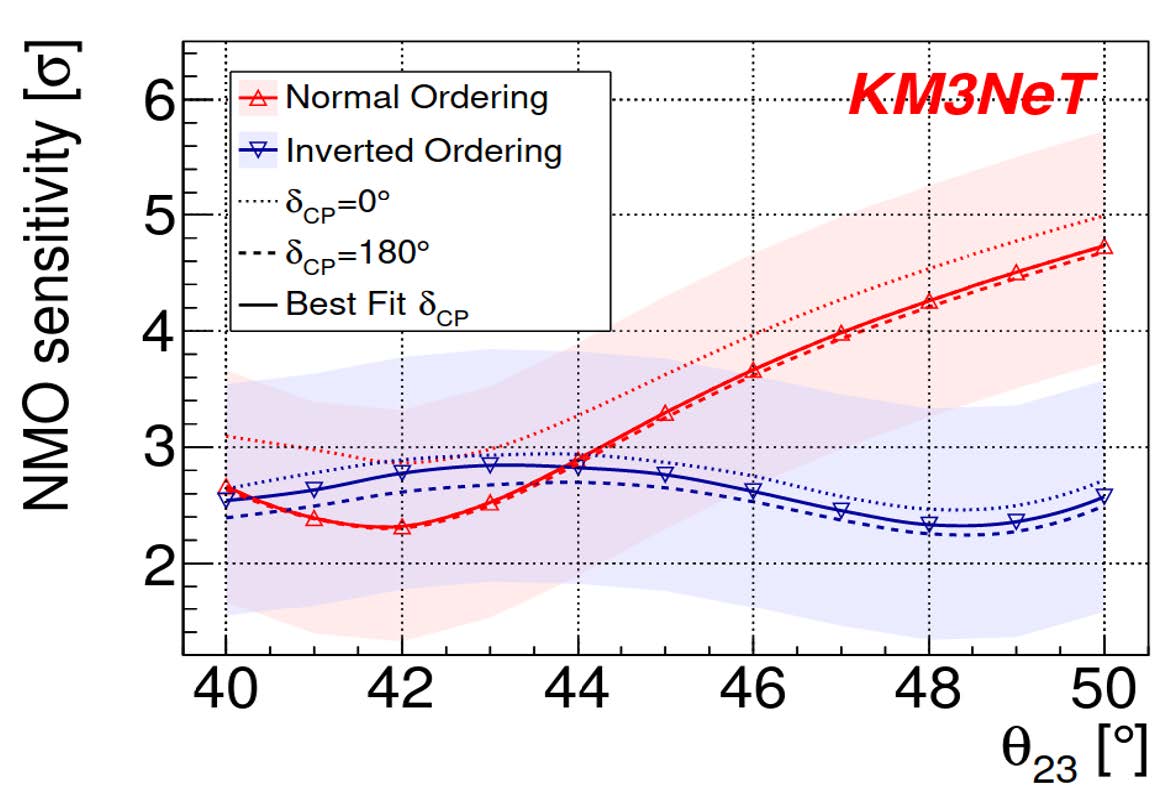}
\caption{Sensitivity of KM3NeT/ORCA to neutrino mass ordering after 3 years of data taking for  normal (red triangles) and inverted ordering (blue triangles) under the hypothesis of three different values of  $\delta_{CP}$.}
\label{fig:nmo}
\end{figure}
\begin{figure}[h]
\includegraphics[width=0.5\textwidth]{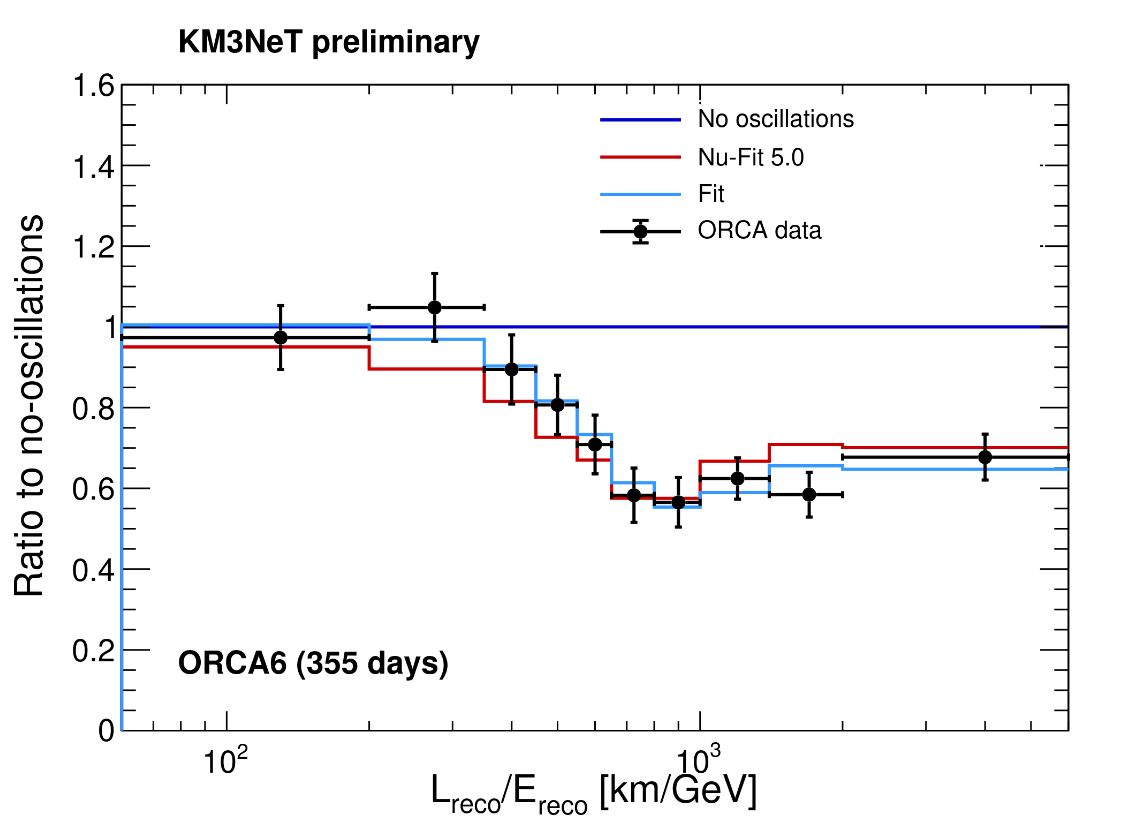}
\hfill
\includegraphics[width=0.5\textwidth]{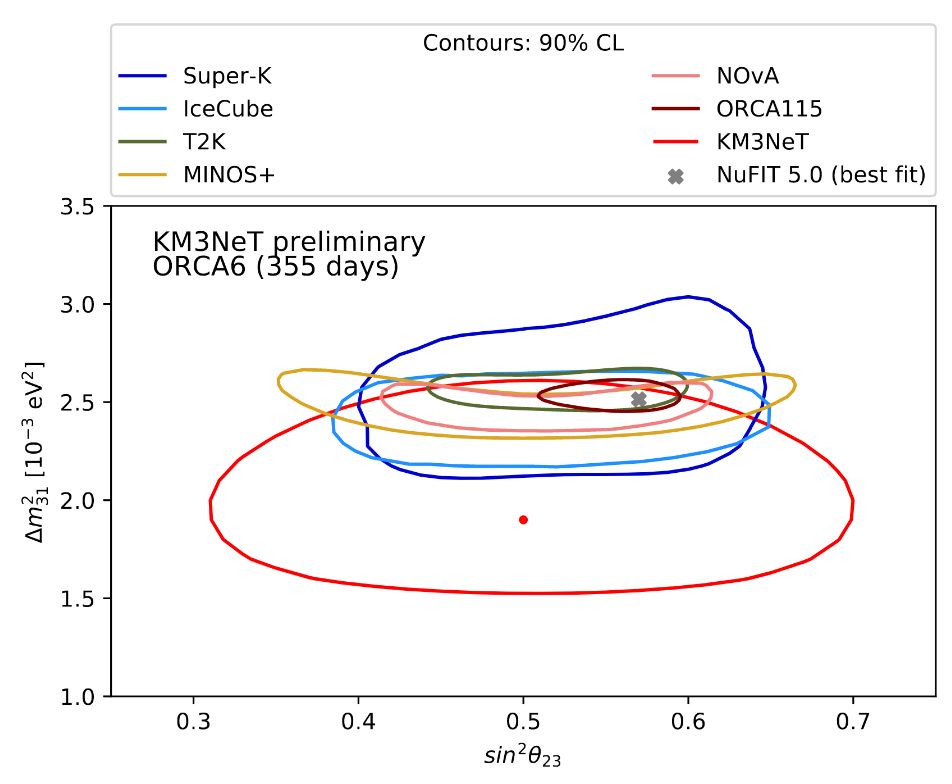}
\caption{Left: Oscillation pattern of atmospheric neutrinos measured with almost one year of data taking with six ORCA DUs compared to no-oscillation hypothesis and to NuFit 5.0 expectation  \cite{nufit}.}
\label{fig:oscill}
\end{figure}

Both ARCA and ORCA are sensitive to the electron antineutrino flux emitted during a core collapse supernovae (CCSN), thanks to the innovative design of the DOM. A large amount of energy is released during a core-collapse supernova event through the emission of neutrino bursts with an average energy  around 10-20 MeV, on a timescale of about ten seconds. These neutrinos cannot individually produce a clear event signature in KM3NeT. Nonetheless a global increase of coincidence counting rate on each single DOM is expected, as shown in Fig.  {\ref{fig:ccsnsensi}}(right). The multiplicity indicated in the figure is the number of PMTs in a DOM hit within a time interval of 10 ns.  The  sensitivity to CCSN events has been investigated for different progenitor masses and distances and an online system for CCSN detection has been implemented, \cite{ccsn}.  Results are shown in Fig. {\ref{fig:ccsnsensi}}(left).

\begin{figure}[h]
\includegraphics[width=0.5\textwidth]{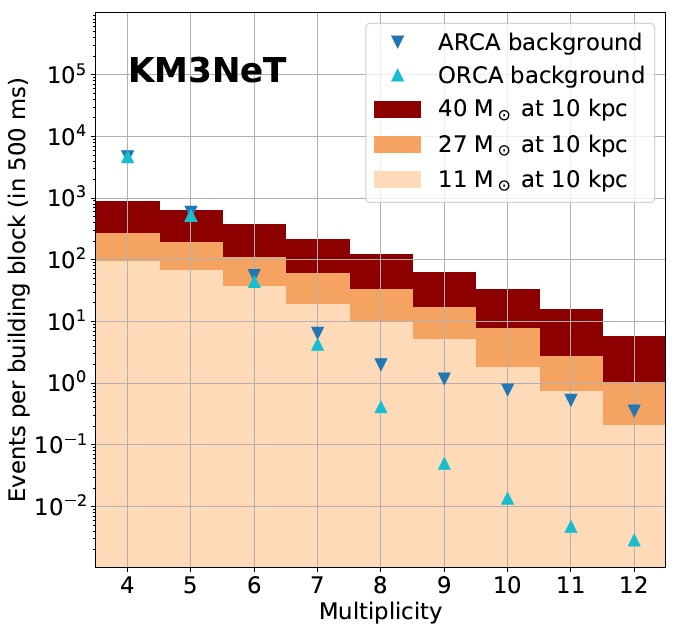}
\hfill
\includegraphics[width=0.6\textwidth]{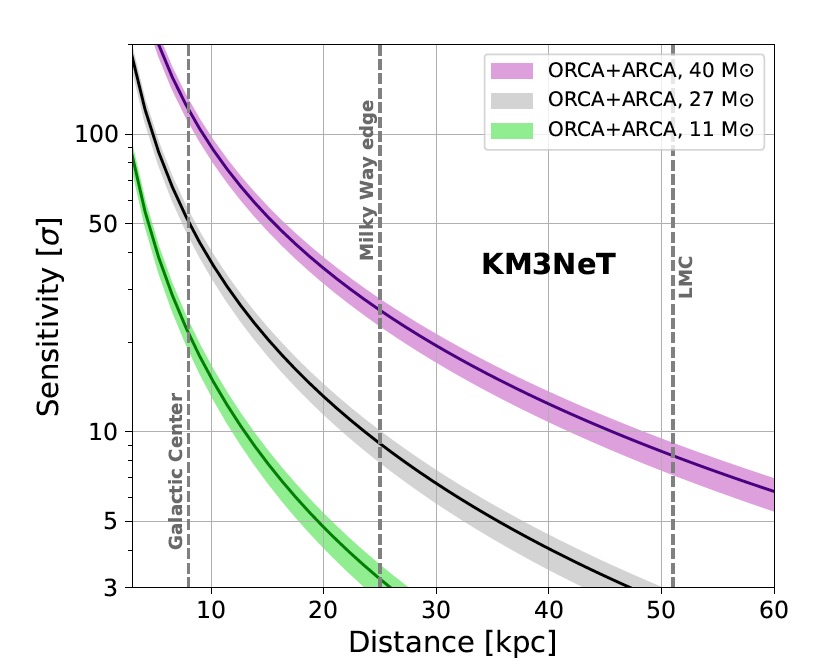}
\caption{Left: number of expected events in a time window of 500 ms for a single BB as a function of the multiplicity (number of PMTs of the same DOM hit within 10 ns). Right: Sensitivity of KM3NeT to the CCSN for three progenitor masses, 11 (green), 27(grey) and 40 (purple) solar masses, as a function of the distance. Error bars include systematic uncertainties coming from models and detector. }
\label{fig:ccsnsensi}
\end{figure}

\section{Conclusion}
After the successful experience of the ANTARES detector (https://antares.in2p3.fr/), which showed the reliability of the Cherenkov technique for neutrino detection, a new generation of undersea neutrino telescopes is being installed at the bottom of  the Mediterranean sea. The KM3NeT infrastructure will host a network of telescopes covering a wide range of neutrino energy, ARCA and ORCA. Thanks to the innovative design of the optical sensor, the Digital Optical Module, the location in the Northern Hemisphere and the excellent optical properties of sea water, KM3NeT/ARCA will be able to explore the Galactic Plane and Centre complementing the field of view of the IceCube detector. The sensitivity of KM3NeT/ORCA will allow to determine the mass ordering of neutrinos in less than six years even in the more pessimistic scenario.

\section*{Acknowledgements}
The authors acknowledge the financial support of the funding agencies:
Agence Nationale de la Recherche (contract ANR-15-CE31-0020), Centre National de la 
Recherche  Scientifique (CNRS), Commission Europ\'eenne (FEDER fund and Marie Curie Program), Institut Universitaire de France (IUF), LabEx UnivEarthS (ANR-10-LABX-0023 and ANR-18-IDEX-0001), Paris \^Ile-de-France Region, France;
Deutsche Forschungsgemeinschaft (DFG), Germany;
The General Secretariat of Research and Technology (GSRT), Greece;
Istituto Nazionale di Fisica Nucleare (INFN), Ministero dell'Universit\`a e della Ricerca (MIUR), PRIN 2017 program (Grant NAT-NET 2017W4HA7S) Italy;
Ministry of Higher Education, Scientific Research and Innovation, Morocco, and the Arab Fund for Economic and Social Development, Kuwait;
Nederlandse organisatie voor Wetenschappelijk Onderzoek (NWO), the Netherlands;
The National Science Centre, Poland (2015/18/E/ST2/00758);
National Authority for Scientific Research (ANCS), Romania;
Ministerio de Ciencia, Innovaci\'{o}n, Investigaci\'{o}n y Universidades (MCIU): Programa Estatal de Generaci\'{o}n de Conocimiento (refs. PGC2018-096663-B-C41, -A-C42, -B-C43, -B-C44 and refs. PID2021-124591NB-C41, -C42, -C43) (MCIU/FEDER, Generalitat Valenciana: Prometeo (PROMETEO/2020/019), Grisol\'{i}a (refs. GRISOLIA/2018/119, /2021/192) and GenT (refs. CIDEGENT/2018/034, /2019/043, /2020/049, /2021/023) programs, Junta de Andaluc\'{i}a (ref. A-FQM-053-UGR18), La Caixa Foundation (ref. LCF/BQ/IN17/11620019), EU: MSC program (ref. 101025085), Spain;

\nolinenumbers

\end{document}